# Two-dimensionality of metallic surface conduction in $Co_3Sn_2S_2$ thin films

Junya Ikeda[1], Kohei Fujiwara[1*], Junichi Shiogai[1], Takeshi Seki[1,2], Kentaro Nomura[1,2], Koki Takanashi[1,2,3], and Atsushi Tsukazaki[1,2,3]

[1] *Institute for Materials Research, Tohoku University, Sendai 980-8577, Japan*

[2] *Center for Spintronics Research Network (CSRN), Tohoku University, Sendai 980-8577, Japan*

[3] *Center for Science and Innovation in Spintronics (CSIS), Core Research Cluster, Tohoku University, Sendai 980-8577, Japan*

[*]Author to whom correspondence should be addressed: kfujiwara@imr.tohoku.ac.jp

**Two-dimensional (2D) surface of the topological materials is an attractive channel for the electrical conduction reflecting the linearly-dispersive electronic bands. By applying a reliable systematic thickness $t$ dependent measurement of sheet conductance, here we elucidate the dimensionality of the electrical conduction paths of a Weyl semimetal $Co_3Sn_2S_2$. Under the ferromagnetic phase, the 2D conduction path clearly emerges in $Co_3Sn_2S_2$ thin films, indicating a formation of the Fermi arcs projected from Weyl nodes. Comparison between 3D conductivity and 2D conductance provides the effective thickness of the surface conducting region being estimated to be approximately 20 nm, which is rather thicker than 5 nm in topological insulator $Bi_2Se_3$. This large value may come from the narrow gap at Weyl point and relatively weak spin-orbit interaction of the $Co_3Sn_2S_2$. The emergent surface conduction will provide a pathway to activate quantum and spintronic transport features stemming from a Weyl node in thin-film-based devices.**

## Introduction

The surface of the films forms interface with substrate or capping layer, which usually plays an active role for the electrical conduction in addition to the bulk region. Various origins for the interface conduction in trivial semiconductors have been explored as charge discontinuity at oxide interface[1], charge transfer[2], dipoles effect[3] and electric polarization effect[4,5]. The effective thickness of the two-dimensional (2D) conductive channel is basically dominated by charge accumulation based on Poisson equation. Besides these interface effects, the emergent topological materials hold intrinsic surface conducting channels that are a Dirac surface state in three-dimensional (3D) topological insulator (TI) (ref. 6) and Fermi arc in Dirac and Weyl semimetals[7,8]. The surface conduction in the topological materials exhibits the intriguing features of linearly-dispersive electronic bands such as spin-momentum locking[9,10], quantum anomalous Hall effect[11,12] and magnetoresistance oscillation at the Weyl orbits[13,14]. As another important feature, the surface states of TIs suddenly disappear by hybridization between two surfaces of top and bottom in ultrathin films. While the dominant factor of the penetration function along thickness direction has been experimentally[15,16] and theoretically[17] investigated, the critical thickness for 3D-TIs is about 5 nm for $Bi_2Se_3$ (ref. 15) and 2 nm for $Bi_2Te_3$ (ref. 16). However, surface conduction in Weyl semimetal (WSM) has been rarely explored in thin films. Recently, a Co-kagome magnet of $Co_3Sn_2S_2$, crystal structure of which is shown in Fig. 1a, has been extensively investigated as a Dirac semimetal (DSM) at paramagnetic phase and a WSM at ferromagnetic phase[18-23]. Large anomalous Hall conductivity has been focused with perpendicular magnetic anisotropy owing to large contribution of Berry curvature in the Weyl nodes[18]. By tilting the symmetric line of the two Weyl nodes from z-axis, surface Fermi arc has been observed by spectroscopies[21-23]. The contribution of the projected Weyl nodes would be detected in electrical conduction for ultrathin films with suppressed bulk conduction though it has not been apparently obtained in bulk $Co_3Sn_2S_2$.

The electrical conduction in thin films at zero magnetic field is generally expected to follow the Ohm's law based on the 3D uniform conductivity $\sigma_{xx}$ ($\Omega^{-1}cm^{-1}$) in whole region of the film; $G = \frac{I}{V} = \sigma_{xx}\frac{tW}{L}$, where $G$ is the conductance of the sample, $I$ current, $V$ applied voltage, $W$ width, $L$ length, and $t$ thickness of the channel. The sheet conductance is experimentally calculated by $G_s = \frac{IL}{VW} = t\sigma_{xx}$. When the surface or interface contributes to the electrical conduction in the films, the sheet conductance of the film $G_{total}^{2D}$ sums up the $t$-dependent conductance in bulk region $G_{bulk} = t\sigma_{bulk}^{3D}$ and the $t$-independent 2D surface conductance $G_{surf}^{2D}$, $G_{total}^{2D} = t\sigma_{bulk}^{3D} + G_{surf}^{2D}$ as shown in Fig. 1b and 1c, where $\sigma_{bulk}^{3D}$ is the 3D conductivity of the bulk region. Experimentally, $t$ dependence of the 2D sheet conductance $G_{\text{total}}^{\text{2D}}$ can separately elucidate the conduction path at 3D bulk and 2D surface. However, it is often difficult to achieve the electrical conduction in the ultrathin layers because the island-like grains initially disconnect each other. This is so-called a dead layer at the interface, which is usually evaluated by the $t$-dependent conductance measurement. To accurately evaluate the bulk conductivity $\sigma_{\text{bulk}}^{\text{3D}}$ of the thin films, therefore, the $t$-dependent measurement is reliable with elimination of the additional conduction at the surface and/or with accurate thickness contributing conduction except for a dead layer. In this study, we applied this analysis to the $Co_3Sn_2S_2$ thin films to elucidate the dimensionality of the conduction channels at bulk and surface separately. The films holding WSM phase are intensively discussed in thickness 23 – 61 nm. The metallic conduction of the films at the WSM phase strongly relates on the superior surface conduction.

**Results**

Temperature dependent resistivity $\rho_{xx}$ of two $Co_3Sn_2S_2$ films with thicknesses of 41 and 40 nm is shown in Fig. 2a with a reference $\rho_{xx}$ of bulk crystal[18]. The two films are typical examples for 12 films with various thicknesses $t$ (23 – 61 nm), prepared by radio-frequency sputtering (Methods) (refs. 24,25). All three traces in Fig. 2a clearly present a kink around 180 K coming from the ferromagnetic transition, corresponding to a phase transition from DSM at $T > T_C$ to WSM at $T < T_C$, where $T_C$ represents the Curie temperature. This feature is in fact detected in magnetization and anomalous Hall conductivity[24,25]. Since metallic conduction is more pronounced in the WSM phase, here we define the residual resistivity ratio ($RRR = \rho_{xx}$ ($T = 300$ K) / $\rho_{xx}$ ($T = 2$ K)) to characterize the metallicity of the films. This $RRR$ is a good measure to classify the films into two groups while the structural qualities of these films are comparable in the inspection by x-ray diffraction[25]. In Fig. 2b, the *RRR* for all the films measured in this study is summarized to classify the films into Groups A (blue) and B (red), in which the border for classification is *RRR* ~ 4. Typical $\rho_{xx}$ values for the film #1 of Group A and the film #2 of Group B with comparable *t* ~ 40 nm represent large and small *RRR* in Fig. 2a, respectively. In comparison with the reference $\rho_{xx}$ of bulk crystal with large *RRR* ~ 8.5, $\rho_{xx}$ of #1 at low temperature is comparable to that of the bulk but the $\rho_{xx}$ of both the films at high temperature is largely different.

Thickness *t* dependence of sheet conductance ($G_s = 1/R_s$, where $R_s$ is sheet resistance) at *T* = 300 K and 2 K is plotted as a function of *t* in Fig. 2c and 2d, respectively. In DSM phase at 300 K, the sheet conductance linearly depends on *t*. In addition, the intercept of the fitting line is close to the origin, holding a dead layer thickness of roughly 1 nm. This linear trend and the small intercept support the validity of this *t*-dependent analysis based on the uniform crystalline quality of the films. The slope in this plot corresponds to the conductivity of the bulk $\sigma_{\text{bulk}}^{\text{3D}}$ as a usually-discussed fundamental physical parameter. In contrast, the intercept of the fitting line at *T* = 2 K in Fig. 2d apparently reveals the emergence of *t*-independent 2D conduction path at extrapolated *t* = 0 nm, corresponding to the 2D surface conductance $G_{\text{surf}}^{\text{2D}}$. This dramatic variation of the intercept in the fitting line from 300 K to 2 K strongly indicates the unique surface transport of the Fermi arc states, relating on the spontaneous symmetry breaking and Fermi energy. The Fermi energy in the non-magnetic phase is far from the Dirac point[26], being the maximum density of state that relates on the Stoner criterion for the ferromagnetism in $Co_3Sn_2S_2$. By contrast, the Fermi energy becomes rather close to the Weyl nodes below $T_C$ (refs. 18,26), which is the origin for the distinct appearance of the surface state in only WSM phase.

At the various temperatures, *t*-dependent analysis was carried out to elucidate the $\sigma_{\text{bulk}}^{3D}$ and $G_{\text{surf}}^{2D}$. In addition, the total conductivity $\sigma_{xx}$ of the films is calculated by conventional scheme $\sigma_{xx} = \frac{\rho_{xx}}{(\rho_{xx}^2 + \rho_{yx}^2)}$, where $\rho_{xx}$ is resistivity and $\rho_{yx}$ Hall resistivity for estimation under the assumption of uniform conduction in whole region. Figure 3a and 3b shows the $\sigma_{xx}$ (black line) for the films #1 and #2 and the analyzed $\sigma_{\text{bulk}}^{3D}$ (red circles) as discussed in Fig. 2c and 2d for the Group A and B, respectively. Above $T_C \sim 180$ K, the analyzed $\sigma_{\text{bulk}}^{3D}$ matches well with the $\sigma_{xx}$, meaning that the analysis is valid for estimation of bulk conductivity for DSM phase. In WSM phase below $T_C$, the $\sigma_{\text{bulk}}^{3D}$ (red closed circles) gradually increases with decreasing temperature, implying the suppression of spin disorder scattering and/or magnon-phonon scattering[19,27-29] as like half-metallic ferromagnetic materials. Moreover, the $\sigma_{xx}$ for #1 (black line) represents superior metallic behavior at low temperature, the value of which is much higher than the $\sigma_{\text{bulk}}^{3D}$. This deviation indicates the contribution of additional conduction path to that of the *t*-dependent bulk, which is the *t*-independent surface conduction (blue region). By contrast, in Fig. 3b, the $\sigma_{xx}$ for #2 (black line) is comparable to the $\sigma_{\text{bulk}}^{3D}$ (red open circles) for Group B. Considering the comparable $\sigma_{\text{bulk}}^{3D}$ for Group A and B, the superior metallic conduction of $\sigma_{xx}$ for #1 is mainly driven by the surface conduction.

The analyzed $G_{\text{surf}}^{2D}$ for Group A and B is plotted as a function of temperature in Fig. 4a. The $G_{\text{surf}}^{2D}$ increases with decreasing temperature in WSM phase for both Groups. The steep increase of $G_{\text{surf}}^{2D}$ for Group A induces the large $\sigma_{xx}$ of the films in Group A. Judging from the consistent trend between the $\sigma_{xx}$ shown in Fig. 3 and the $G_{\text{surf}}^{2D}$ in Fig. 4a, the difference between the observed $\sigma_{xx}$ (black line) and the analyzed $\sigma_{\text{bulk}}^{3D}$ (red circles) in Fig. 3; $\sigma_{xx} - \sigma_{\text{bulk}}^{3D} = \sigma_{\text{surf}}^{3D}$, corresponds to the contribution of 2D $G_{\text{surf}}^{2D}$. Based on this assumption, we estimate the effective thickness of the surface conduction $t_{\text{surf}}^{\text{eff}}$ by a following equation

$\sigma_{\text{surf}}^{\text{3D}} \times t_{\text{surf}}^{\text{eff}} \times 2 = G_{\text{surf}}^{\text{2D}}$ where the factor of 2 corresponds to two surfaces of the films. Although the error bar is large in Fig. 4b, the *t*-independent 2D conduction region is estimated to be roughly 20 nm constant against *t*-variation. This value is rather thicker than that of 3D-TI for example 5 nm of $Bi_2Se_3$ (ref. 15). The distribution of surface Fermi arc along *z*-direction may be related to the broadness of the projection of Weyl nodes with narrow gap[17]. In addition, the weaker spin-orbit coupling of $Co_3Sn_2S_2$ compared to 3D-TIs such as $Bi_2Se_3$ and $Bi_2Te_3$ may induce the weak confinement of surface state, resulting in the large effective thickness (refs. 15,17). While we here assume uniform $G_{\text{surf}}^{\text{2D}}$ to estimate the averaged effective thickness, the penetration function of the surface conduction region should be more carefully considered to be rapid decay along *z*-direction as like 3D-TIs (refs. 17,30). In this study, apparent detection of the *t*-independent conductance reveals the possibility of surface Fermi arc conduction in the $Co_3Sn_2S_2$ thin films.

**Discussion**

Here we discuss the origin for the appearance of *t*-independent surface conduction in the films. The robust *t*-independent current path is experimentally detected by the analysis of sheet conductance of the films. The thickness of the films is enough thick to form uniform crystalline structure and to hold a feature of WSM with large anomalous Hall conductivity[25]. The top and bottom surfaces of the films form the interface with a capping layer of $SiO_x$ and $Al_2O_3$ substrate, respectively. Such interface between oxide insulator and sulfide metal is likely to be basically inactive for the electrical conduction. Other extrinsic defect formation contributing to the metallic electrical conduction can be excluded. As intrinsic origins, the surface projected Fermi arc is possibly located at the inert interface. Moreover, the appearance of surface conduction below $T_C$ should be noticeable to the Weyl nodes stemming from WSM phase. However, the origin for different values of *RRR* in two groups is not obvious from the structural quality. In fact, the analyzed value of $\sigma_{\text{bulk}}^{\text{3D}}$ for both group is comparable. As shown in Fig. 2b, most of the films in Group A are roughly thicker than those in Group B, implying that 40 nm ~ 2 × $t_{\text{surf}}^{\text{eff}}$ looks criteria for the large surface conduction. The surface conduction in the WSMs may be weakened by hybridization each other. Single crystalline bulk also presents large *RRR* (ref. 18), even though the electrical conduction is dominated by the $G_{\text{bulk}}$ due to thick limit. To be consistent with the bulk results, the detected surface conduction is likely to shed light on the thin films with less contribution of bulk conduction. In addition, the surface termination might be a plausible origin for holding different Fermi surface contributing to the electrical conduction[7,15,21,31], although determination of the surface termination of $Co_3Sn_2S_2$ thin film is difficult in the present sample structure. The superior metallic surface conduction in Group A may link to the rich surface Fermi arc by sulfur termination. By contrast, the less surface conduction in Group B may be terminated by Sn. The further studies on specific interface formation need to clarify the termination and contribution to the charge distribution.

**Summary and conclusions**

Systematic thickness dependence of conductance reveals the emergence of *t*-independent 2D conduction path in the $Co_3Sn_2S_2$ films under Weyl semimetal phase. Based on the analysis for two group films, superior metallic conduction at surface dominates the large *RRR* of the films. Origin for the appearance of 2D metallic conduction path is likely the surface Fermi arc projected from Weyl nodes. Broad penetration of the 2D surface state comes from material specific parameters of Weyl semimetal $Co_3Sn_2S_2$ such as strength of spin-orbit interaction and gap size at Weyl point. The interface formation with the $Co_3Sn_2S_2$ will provide an attractive arena for quantum transport and spintronic phenomena based on the intriguing 2D metallic surface state.

**Materials and Methods**

**Film preparation**

Co$_3$Sn$_2$S$_2$ thin films were prepared by radio-frequency sputtering. The growth procedure is following and detail information can be found in previous studies[24,25]. Firstly, a Co$_3$Sn$_2$S$_2$ film was deposited on Al$_2$O$_3$ (0001) substrate at 400 °C. Then, a SiO$_x$ capping layer was deposited on the Co$_3$Sn$_2$S$_2$ film. The film was annealed at 800 °C in vacuum for 1 hour. The crystal structures and composition were characterized by x-ray diffraction and energy-dispersive x-ray spectroscopy. The $t$ (23 – 61nm) was controlled by the deposition duration, which was confirmed by the Laue fringe in x-ray diffraction patterns.

**Electrical transport measurement**

Electrical measurements were carried out in physical properties measurements system (PPMS, Quantum Design) (ref. 25). The films were scratched to make Hall-bar shape with indium contact electrodes. In the previous studies, the large anomalous Hall conductivity was observed in all films, indicating that the WSM phase is materialized.

**Acknowledgements**     The authors are grateful to K. Kobayashi, Y. Yanagi, M.-T. Suzuki, and Y. Motome for fruitful discussions. This work was performed under the Inter-University Cooperative Research Program of the Institute for Materials Research, Tohoku University (Proposal No. 19G0410). This work was supported by JSPS KAKENHI (Grant No. 20H01830) and JST CREST (JPMJCR18T2).

**Author Contributions**     J.I. and K.F. grew the films and measured the electrical transport properties. J.S. performed the magnetoresistance measurements. K.F. performed the magnetization measurements under the support by J.S., T.S., and K.T. J.I. and K.F. analyzed the measured data. K.N. contributed to theoretical interpretations of the experimental results. J.I., K.F., and A.T. wrote the manuscript with input from other authors. All authors discussed the results. A.T. supervised the project.

**Additional information**     Reprints and permissions information is available online at www.nature.com/reprints. Correspondence and requests for materials should be addressed to K.F.

**Competing financial interests**     The authors declare no competing financial interests.

**Figure 1 | Electrical conduction in Weyl semimetal $Co_3Sn_2S_2$ thin films. a**, Crystal structure of $Co_3Sn_2S_2$. **b**, Concept of electrical conduction of $Co_3Sn_2S_2$ thin films in the bulk $G_{bulk}$ and at the surface $G_{surf}$. Below Curie temperature, perpendicular magnetization $M$ (green arrows) induces anomalous Hall effect (AHE). **c**, Cross-section of thin films with thickness $t$-dependent $G_{bulk}^{2D}$ and $t$-independent $G_{surf}^{2D}$. Sheet conductance in the bulk $G_{bulk}^{2D}$ is evaluated by three-dimensional conductivity of bulk $Co_3Sn_2S_2$ $\sigma_{bulk}^{3D} \times t$.

**Figure 2 | Temperature $T$ and thickness $t$ dependence of electrical conduction in $Co_3Sn_2S_2$ thin films. a,** $T$ dependence of resistivity $\rho_{xx}$ for a reference of bulk (black curve from ref. 18), film #1 (blue curve; $t$ = 41 nm) from Group A, and film #2 (red curve; $t$ = 40 nm) from Group B. **b,** Residual resistivity ratio $RRR$ for the films as a function of thickness, which guided to categorize the films to Group A and B by the criteria at roughly $RRR$ = 4. **c** and **d**, To detect the $t$-independent conductance, sheet conductance $G_{total}^{2D}$ for various thickness films at $T$ = 300 K and 2 K are plotted, respectively. The slope corresponds to three dimensional conductivity $\sigma_{bulk}^{3D}$. The detection of intercept in d indicates the emergence of $t$-independent two-dimensional surface conductance $G_{surf}^{2D}$. The experimental data in a and d are replotted from ref. 25.

**Figure 3 | Analyzed conductance for bulk and surface in $Co_3Sn_2S_2$ thin films. a** and **b,** The $\sigma_{bulk}^{3D}$ is estimated by the slope in $t$-dependent sheet conductance for Group A and B as shown in Fig. 2c and d. Temperature dependence of the analyzed $\sigma_{bulk}^{3D}$ is plotted with the obtained conductance data for the representative film #1 in A and #2 in B (black line). The deviation between the black line and $\sigma_{bulk}^{3D}$ corresponds to the contribution of electrical conduction at the surface $\sigma_{surf}^{3D}$.

**Figure 4 | Comparison of 2D conduction $G_{surf}^{2D}$ and averaged 3D thickness of the surface conduction path. a,** Temperature dependence of the surface conduction $G_{surf}^{2D}$ by the analysis as shown in Fig. 2d for the data of Group A (blue) and B (red). **b,** Under the assumption of uniform electrical conduction at the surface region, the effective thickness of surface conduction $t_{surf}^{eff}$ is calculated by $t_{surf}^{eff} = \frac{G_{surf}^{2D}}{2\sigma_{surf}^{3D}}$ at $T = 2$ K.

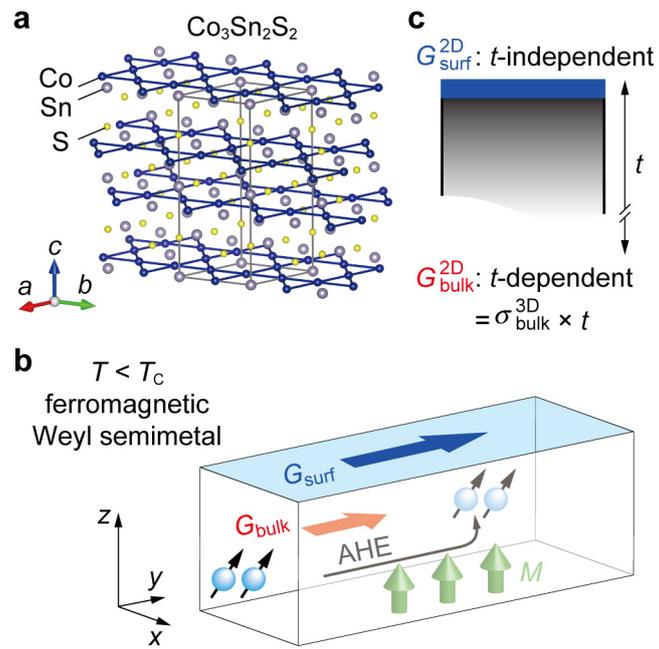

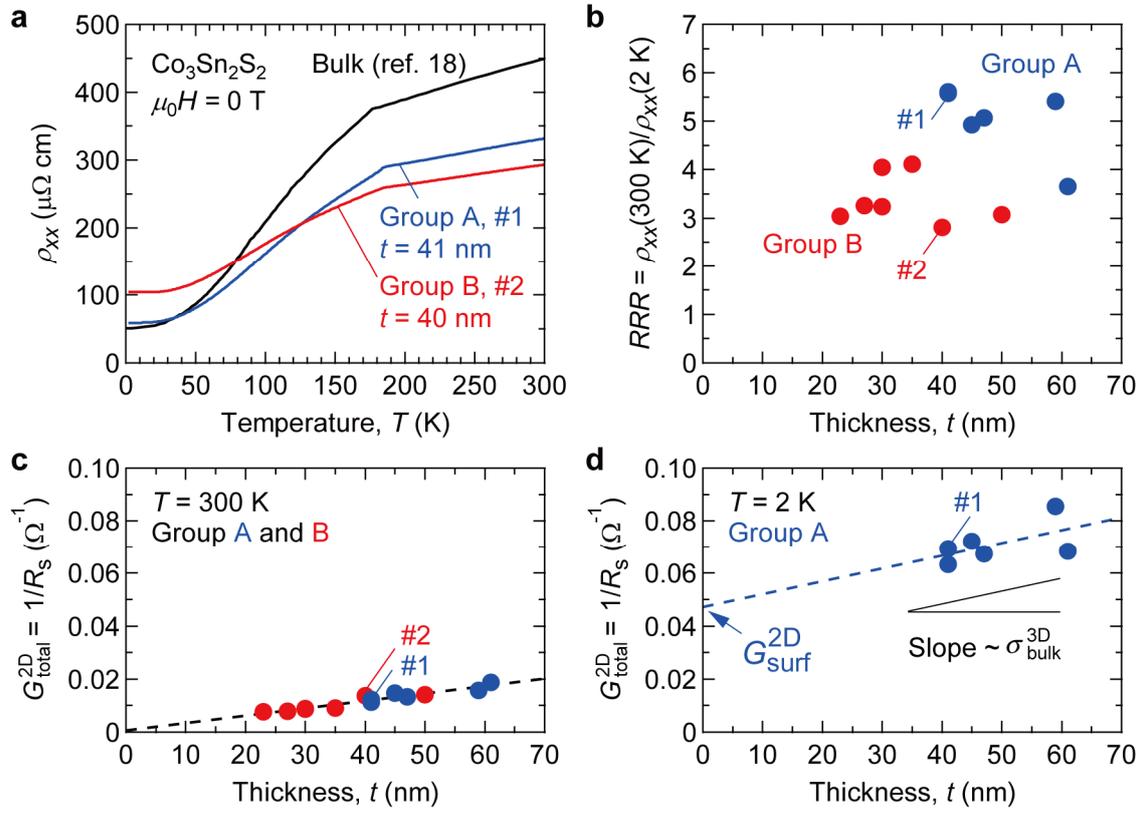

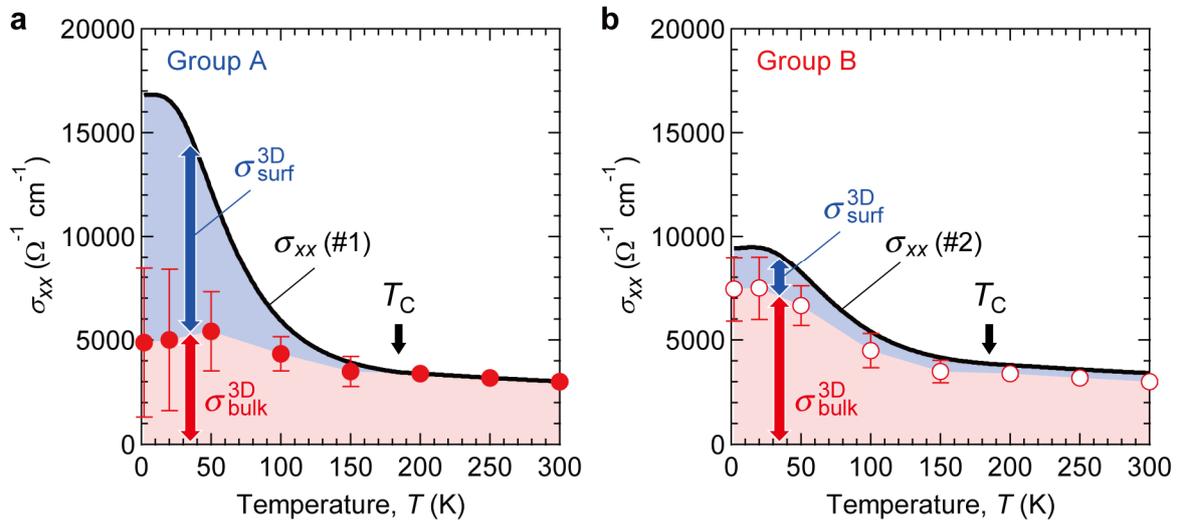

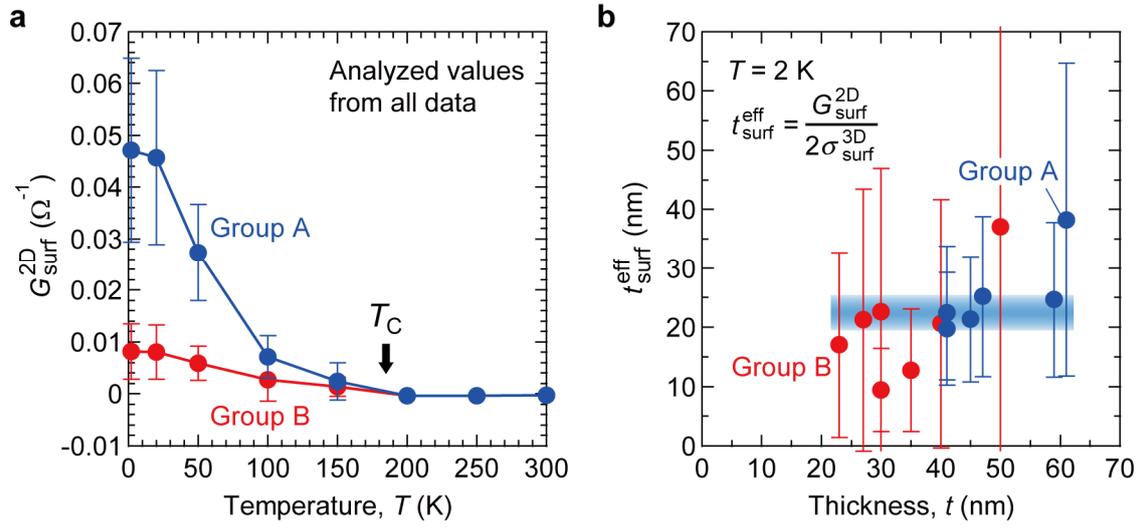

J. Ikeda *et al*., Figure 4